\documentclass[sigconf,nonacm,balance]{acmart}
\usepackage{xspace}
\usepackage[inline]{enumitem}
\usepackage{amsmath}
\usepackage{cleveref}
\usepackage{csquotes}
\usepackage{fontawesome}
\usepackage{amsmath}
\usepackage{tikz}
\usetikzlibrary{positioning, arrows, arrows.meta, calc, patterns, positioning}
\newcommand\icon[1]{{\large#1}}
\newcommand\badge[3]{\icon{#1}\hspace{-.5em}\raisebox{-.4em}{\color{#2}\footnotesize#3}}

\newcommand\block[5]{
    \node[block, fill=#1!15, draw=#1!40, #2] (#3) {};
    \node[yshift=.3cm] at (#3.south) {#5};
    \node[yshift=.7cm] at (#3.south) {{\LARGE#4}};
}

\setcopyright{none}
\newcommand{\eg}{e.g.,\xspace}

\newcommand{\ie}{i.e.,\xspace}

\newcommand{\WAIT}{\textsc{WAIT}\xspace}
\newcommand{\WAITlong}{Web Application Integrity Transparency\xspace}

\definecolor{uulm}{RGB}{125,154,170}
\definecolor{uulm-akzent}{RGB}{169,162,141}
\definecolor{uulm-in}{RGB}{163,38,56}
\definecolor{uulm-med}{RGB}{38,84,124}
\definecolor{uulm-mawi}{RGB}{86,170,28}
\definecolor{uulm-nawi}{RGB}{189,96,5}

\begin{document}
\title[WAIT: Web Application Integrity Transparency]{\texorpdfstring{WAIT: Protecting the Integrity of Web Applications\\ with Binary-Equivalent Transparency}{WAIT: Protecting the Integrity of Web Applications with Binary-Equivalent Transparency}}

\author{Dominik Mei{\ss}ner}
\orcid{0000-0002-2937-6306}
\email{dominik.meissner@uni-ulm.de}
\affiliation{%
  \department{Institute of Distributed Systems}
  \institution{Ulm University, Germany}
  \streetaddress{Albert-Einstein-Allee 11}
  \postcode{89081}
  \city{}
  \country{}
}

\author{Frank Kargl}
\orcid{0000-0003-3800-8369}
\email{frank.kargl@uni-ulm.de}
\affiliation{%
  \department{Institute of Distributed Systems}
  \institution{Ulm University, Germany}
  \streetaddress{Albert-Einstein-Allee 11}
  \postcode{89081}
  \city{}
  \country{}
}

\author{Benjamin Erb}
\orcid{0000-0002-5432-4989}
\email{benjamin.erb@uni-ulm.de}
\affiliation{%
  \department{Institute of Distributed Systems}
  \institution{Ulm University, Germany}
  \streetaddress{Albert-Einstein-Allee 11}
  \postcode{89081}
  \city{}
  \country{}
}

\begin{abstract}
Modern single page web applications require client-side executions of application logic, including critical functionality such as client-side cryptography.
Existing mechanisms such as TLS and Subresource Integrity secure the communication and provide external resource integrity.
However, the browser is unaware of modifications to the client-side application as provided by the server and the user remains vulnerable against malicious modifications carried out on the server side.
Our solution makes such modifications transparent and empowers the browser to validate the integrity of a web application based on a publicly verifiable log.
Our \WAITlong (\WAIT) approach requires
\begin{enumerate*}[label=(\arabic*)]
\item an extension for browsers for local integrity validations,
\item a custom HTTP header for web servers that host the application, and
\item public log servers that serve the verifiable logs
\end{enumerate*}.
With \WAIT, the browser can disallow the execution of undisclosed application changes.
Also, web application providers cannot dispute their authorship for published modifications anymore.
Although our approach cannot prevent every conceivable attack on client-side web application integrity, it introduces a novel sense of transparency for users and an increased level of accountability for application providers particularly effective against targeted insider attacks.
\end{abstract}

\maketitle

\section{Introduction}
\label{sec:intro}
An increasing number of modern web applications moves at least parts of their application logic onto the browser.
Rich clients enable more responsive and feature-packed user interfaces and allow the execution of complex application features directly in the browser.
In addition, such applications can emulate the experience of running a native application within the browser\,---\,both on desktop and mobile.
This trend is also reflected in recent developments such as single page applications and progressive web applications.

So far, from the point of view of application providers, the browser has been considered to be an untrusted execution environment.
Malicious users or tampered browsers are able to modify client-side application code.
This is why most client-side functionalities are still backed by server-side validations and client input is generally considered to be tainted at the server side.

With the simultaneous availability of secure communication (\ie Transport Layer Security~\cite{rfc5246}) and cryptographic capabilities in the browser (\ie Web Cryptography API~\cite{W3CryptoAPI}), several web applications have even started to move cryptographic application functionalities away from the server and onto the client side.
This change enables web applications in which sensitive content is encrypted and decrypted only in the browsers, but becomes opaque for the server-side web application\,---\,an interesting property in terms of security and privacy.
As a consequence, an alternative perspective on trust has emerged in which the users are questioning the authenticity of the application code that has been served by the application provider.
More precisely, users might wonder whether the application provider is actually serving the code it promised to serve.
From the users' perspective, the browser now represents a trustworthy execution environment, however the integrity of the client code is challenged.
A prime example for such applications are web-based clients of messaging services that guarantee end-to-end encryption (\eg WhatsApp Web, ProtonMail).
Despite initial trust in the web application, the user cannot track covert or unintended application changes.
Following this example, changes in the client-side logic could break the end-to-end encryption and leak messages in plain text.
Maliciously modifying the client-side application code and selectively serving it to certain clients represents a covert and targeted attack against individuals.
Such an attack is likely to occur by malicious insiders at the application provider, either for personal gains or for instance on behalf of state actors.
As the victim cannot easily compare the client-side application code it has received from the server to the code other users have received, it is very difficult to detect such attacks.
In other words, the user cannot authenticate the client-side application code against an authoritative source.

Existing mechanisms for increasing security and integrity of web content cannot address this issue.
Transport Layer Security (TLS)~\cite{rfc5246} prevents the modification of web resources during transmission.
However, if the integrity of web resources is already compromised at the server, TLS provides no further protection.
Content Security Policy (CSP)~\cite{W3CSP} is an additional security layer of browsers to mitigate cross site scripting.
Again CSP does not cover this, as we consider malicious code to be served by an legitimate server.
Subresource Integrity (SRI)~\cite{W3SRI} is a security mechanism to detect unexpected modifications of external resources (\eg resources from a content delivery network).
Due to the issue of bootstrapping trust, SRI cannot be used by a web application to prove its own integrity.
What is missing is a concept that provides integrity validation for web applications in the browser on behalf of the user.

The required concept must seamlessly integrate into existing web technology stacks and should ideally build on established security and web mechanisms.
In addition, the concept should be lightweight to minimize adverse effects on performance.
Once deployed, it should provide an opt-in protection for web applications and users with heightened security or privacy requirements.
When enabled, the protection mechanism should locally block browser access to web applications that have unpublished modifications.

We suggest \WAIT (\WAITlong)\,---\,a security concept that hinders specific attacks on the integrity of web applications.
In addition, we release a proof-of-concept prototype implementation and evaluate its performance impacts.

\section{Related Work}
\label{sec:relatedwork}
While traditional mechanisms for web content security and integrity do not solve our overall problem, auditable transparency logs share a similar idea.
Conventionally, TLS has the issue that trusted certificate authorities are able to issue certificates for any hostname.
This allows the issuance of fraudulent certificates for arbitrary domains\,---\,either by mistake or with malicious intends\,---\,and without the knowledge of the actual domain owner.
Certificate transparency (CT)~\cite{RFC6962} is a framework to monitor and audit certificates that are issued by certificate authorities.
CT introduces an append-only transparency log based on a Merkle tree, where certificate authorities can submit newly issued certificates to.
While CT does not prevent the use of fraudulent certificates in the first place, it does allow providers to monitor the transparency log for newly issued certificates for their domain names.
Providers can then detect fraudulent certificates in a timely manner.

Binary transparency is transparency approach based on CT introduced by Mozilla that targets the distribution of software executables~\cite{MozillaBinaryTransparency}.
Cryptographic checksums are computed for all binaries of a release and arranged as leave nodes of a Merkle tree.
A TLS certificate is obtained for a release-specific domain name that includes the version and the Merkle tree head.
Software updater, can verify this certificate\,---\,and thus the release\,---\,through a lookup in the CT log.
Contour~\cite{al2018contour} can be seen as a generalized approach towards binary transparency.
Countour utilizes a distributed ledger, the Bitcoin network, as its transparency log.
To efficiently store release information in a Bitcoin transaction, Contour introduces another entity, the so called authority.
It collects a set amount of binaries and\,---\,similar to Mozilla's approach\,---\,constructs a Merkle tree that contains these binaries as leave nodes.
The tree's root is then included in a Bitcoin transaction.
Our goal has many similarities with binary transparency, although the execution environments (\ie native applications vs. client-side JavaScript) and the potential attack targets (\ie modified source code built into application binaries vs. manipulated client-side application code to be delivered to the client) differ.
Eventually, we want to provide a similar level of authenticity for single page web application logic on the client side, as binary transparency provides for binary releases\,---\,once a malicious actor has succeeded in tampering with source code, it cannot be undetectably deployed to users.

\section{The \WAIT Approach}
\label{sec:concept}

Next, we consider relevant parties, define the threat model, and describe the procedures performed by these parties.

\begin{figure}[ht]
  \centering
  \vspace{-.1em}
\begin{tikzpicture}[
    block/.style={
        fill=uulm-med!15,
        draw=uulm-med!40,
        line width=1pt,
        rounded corners,
        inner sep=.25cm,
        text width=3.3cm,
        minimum height=1.0cm,
        inner sep=0,
        align=center
    },
    arrow/.style={
        -latex,
        line width=1.5pt,
        draw=black!70
    }
]

    \block{uulm-mawi}{text width=2.5cm, xshift=.16cm, yshift=.16cm}{log}{\faListAlt}{Transparency Log}
    \block{uulm-mawi}{text width=2.5cm, xshift=.08cm, yshift=.08cm}{log}{\faListAlt}{Transparency Log}
    \block{uulm-mawi}{text width=2.5cm}{log}{\faListAlt}{Transparency Log}
    \block{uulm-med}{right=3cm of log, text width=2.3cm}{developer}{\faUser}{Developer}
    \block{uulm-nawi}{below=.7cm of developer, text width=2.5cm}{server}{\faServer}{Web Server}
    \block{uulm-in}{left=3cm of server, text width=2.5cm}{browser}{\faFirefox}{Web Browser}

    \draw[arrow] ($ (log.east) + (.15, -.1) $) -- node[midway,below,sloped] {\scriptsize\textcolor{uulm-in}{\faCertificate} Promise (PoI) (ii)} ($ (developer.west) + (0, -.1) $);
    \draw[arrow] ($ (developer.west) + (0, 0.1) $) -- node[midway,above,sloped] {\scriptsize\icon{\faFileCodeO} Submit/Update (i)} ($ (log.east) + (.15, 0.1) $);
    \draw[arrow, dash pattern=on 2.5pt off 9pt] ($ (developer.south) + (0, 0) $) -- node[midway,yshift=3pt] {\scriptsize\badge{\faFileCodeO}{uulm-in}{\faCertificate} Deployment (iii)} ($ (server.north) + (0, 0) $);
    \draw[arrow] ($ (server.west) + (0, 0) $) -- node[midway,above] {\scriptsize\badge{\faFileCodeO}{uulm-in}{\faCertificate} App Bundle + PoI (iv)} ($ (browser.east) + (0, 0) $);
\end{tikzpicture}

  \vspace{-.1em}
  \caption{The \WAIT approach.}
  \label{fig:approach}
  \vspace{-.5em}
\end{figure}

\subsection{Involved Parties}
We consider four relevant parties that participate in \WAIT.

\begin{description}[leftmargin=8pt]
    \item[Application Developers/Providers] %
        develop and deploy web applications to provide a service, often with monetary incentives.
        Developers and providers want a large and growing user base and thus require a good reputation as well as trust by users.
        However, individual insiders may be motivated or persuaded to launch a (targeted) attack on the user base, out of personal interest or due to outside intervention, such as state actors or bribe offerings.

    \item[Transparency Log Operators] %
        are independent entities that provide public append-only transparency logs to which applications can be submitted.
        The necessity of trust in log operators can be reduced by cryptographically assuring the append-only properties through appropriate mechanisms, such as Merkle trees.

    \item[Independent Auditors] %
        such as researchers and other security professionals, analyze available web applications by looking at their client-side application code.
        This includes checking the public log entries of an application and auditing a specific version to uncover bugs, vulnerabilities, or deliberate backdoors.

    \item[Service Users] %
        use the web applications and want to minimize privacy risks and keep their data secure.
        Therefore, they want to be assured that they are using a secure and uncompromised version of the application.
        Optionally, they want to be sure to use the latest version of a web application.
\end{description}

\subsection{Threat Model}
\WAIT targets a very specific threat to users, where they are provided with an compromised version of a single page web application.
The compromised release attacks the users' privacy or entails other threats for the users.
We assume attackers to have access to the code that will later be delivered to and executed by the client.
This malicious manipulation of client-side application logic can happen at an arbitrary step of the development cycle, during the build process, or within the deployment pipeline.
Our primary attack vector with this scope are insider attacks in which a certain user falls victim to a targeted attack by an insider.
To a certain extent, we also consider the possibility of entire application providers that are going rogue and break previous agreements.
Here we cannot offer the user any direct protection, because the provider of the application can still announce a legitimate update.
However, the log can at least provide non-repudiation, so the provider has to claim responsibility at a later point in time.
An orthogonal threat that we address separately is a downgrade attack.
Here, an attacker tries to defer an upgrade in order to benefit from a previous release (\eg existing vulnerability).
Other attacks against web server and web application are considered out of scope and have to be addressed by complementing security mechanisms.

\subsection{\WAIT Procedures}
Next, we take a look at the different procedures taking place in the \WAIT approach (also see \Cref{fig:approach}).
As a reminder, we do not target arbitrary web applications in which pages are generated dynamically and individually upon request.
Instead, we only cover single page web applications with a static client-side application logic.
Dynamicity, user interactions, and loading additional, dynamic content (but no scripts) from the server is then handled by single page application via dispatching AJAX or fetch calls.

\subsubsection{Computing the digest of a web application}
Instead of computing a digest over the entire application bundle, we establish transitive trust from the main document of the web application to all resources that comprise the bundle.
We achieve this by combining SRI~\cite{W3SRI} and CSP~\cite{W3CSP} in a novel way and complement them with additional restrictions by extending browsers.
Browsers should only interpret scripts and styles that are statically included in the application bundle.
Code that is dynamically generated or loaded through an XHR request, must not be executed.
By using a CSP that contains a \texttt{default-src} or \texttt{script-src} directive, the browser will not interpret inline scripts and styles and prevent dynamic code evaluation through the \texttt{eval} function or the \texttt{Function} constructor.
We combine such a strict CSP with the requirement that all referenced scripts and styles have an SRI integrity checksum, even if they originate from the same origin.
This combination of CSP and SRI ensures that client side logic is loaded initially and cannot be changed without updating the main document.
Hence, a cryptographically secure digest of the main document can transitively authenticate the entire application bundle.

\subsubsection{Submitting a web application release}
Developers have to submit their application releases to transparency logs before deploying it to their userbase.
This process can partially be integrated within an automated build process of the application, by automatically computing SRI checksums for all resources that are part of the application bundle and referenced by the main document.
An additional meta tag in the main document links the application to the developer identity in the form of the public component of a developer key pair.
To submit the application to a log server, the developer computes the cryptographic hash of the document, sign it with their private key, and upload it to one or more log servers.
Transparency logs verify that the signature matches the application, append the entry to the log, and respond with a cryptographically signed promise to include the entry in the log.
The application's web server has to be configured to include these inclusion promises in responses as a custom \texttt{X-WAIT-Inclusion-Promise} HTTP header.
To mitigate deferred upgrade attacks, a log server may add an expiration date to inclusion promises and prevent developers from submitting a new release before the old one has expired.
Thereby, guaranteeing that there is only one valid version of an application at every point in time.
Such expiration dates should feature a reasonable tolerance to mitigate slightly delayed renewals and small differences due to clock synchronization.
Optionally, the log provider can also act as an archive of application versions, by storing the full application bundles alongside the transparency log.
This simplifies the work of independent auditors, as they can retrieve past versions from archives.
However, we argue that even if an archived version is not available, reputable application developers will provide auditors access to past releases in order to maintain their reputation.

\subsubsection{Updating an inclusion promise}
Developer can resubmit a release to renew an inclusion promise.
A timestamp is included in the update request to provide freshness and to prevent replaying a previous submission.
The log server can check that this request is already included in the append-only log and thus skip the expensive append operation and respond with a fresh inclusion promise.

\subsubsection{Client-side validation}
We propose a complementary security mechanism for web browsers that has to be provided either as an extension or, eventually, as a native component of the browser.
After fetching the application and before executing it, the web browser can verify the transitive trust of any application supporting \WAIT, by
\begin{enumerate*}[label=(\arabic*)]
\item checking for a sufficiently strict CSP,
\item requiring SRI checksums for all resources, and
\item asserting a secure context
\end{enumerate*}.
This check is implemented for all requested resources, except responses to AJAX/fetch requests.
Besides checking the strictness of the CSP and the existence of SRI checksums, the browser also validates the signature of the inclusion promises against a trusted set of transparency logs and that the digest matches the main document.
The trusted set of logs is provisioned with the extension or the browser software, in a similar fashion to the trust store of certificate authorities (\eg as defined by the browser vendor).
The browser may only execute the application if sufficient inclusion promises are provided that originate from a trusted log server as well as match the checksum and URL of the document.
Otherwise, an error message is presented to the user that indicates the trust issue, similar to TLS certificate warnings.
To prevent downgrade attacks by removing the header from a resource that has previously been secured, the extension keeps track of secured resources (similar to HTTP Strict Transport Security~\cite{rfc6797}).

\section{Prototype \& Evaluation}
\label{sec:prototype}

To prove the feasibility of \WAIT and to evaluate its performance overhead in practice, we have implemented a system prototype as well as a demo web application.
We have released prototype implementations for all components required for \WAIT as well as all evaluation artifacts and scripts to reproduce our results\footnote{\url{https://github.com/vs-uulm/wait-prototype}} under an open source license.

\subsection{Proof-of-Concept Implementation}
For the client-side verification, we implemented a custom \emph{browser extension} using the cross-browser WebExtensions API.
For the web development part, we have extended an existing webpack-based \emph{build process} to provide transitive trust for the application bundle, by adding automatically adding SRI checksums.
We implemented a mockup \emph{log server} that exposes an HTTP-based API and simple CLI application to interface with it.
The log service allows to register applications and to update inclusion proofs.
Furthermore, we use an unmodified \emph{web server} (NGINX) which is configured to additionally inject the inclusion proof header for the application path.

\subsection{Prototype Evaluation}
The evaluation measures the initial loading time of a single page application in a browser and consisted of 1,000 individual and isolated loads, without our extension (baseline) and with our extension enabled (\WAIT).
The demo application is a simple note taking application, which uses the browser-native WebCryptography API~\cite{W3CryptoAPI} to locally encrypt a dictionary of private notes in the browser and backup it to a server.
The application consists of 9 web resources with a total size of 1.5 MB.
In each run, we measured the page loading times for the full application page via timestamping (\texttt{performance.timing} API) and with caching disabled.
The evaluation was conducted on a desktop computer with an Intel Core i7-7700 (quad-core with SMT; 3.60 GHz) CPU and 32 GB RAM, running Ubuntu 20.04 LTS with Firefox Nightly 83.0a1.
Loading times increased consistently when \WAIT was enabled ($M_{base}$ = 415 ms, $SD_{base}$ = 82.2 ms vs. $M_{\WAIT}$ = 702 ms, $SD_{\WAIT}$ = 120.0 ms).
On average, loading times of the demo application increased by 69\% when \WAIT was in use due to the local integrity check.
While this represents a pronounced increase in the loading times, we argue that this increase is still reasonable for the targeted web applications\,---\,single-page applications and progressive web applications.
In fact, the integrity check only slows down the initial load, but not the subsequent user interactions with the web application.

\section{Discussion}
\label{sec:discussion}
A major limitation of \WAIT is the narrow focus on those single page web applications which provide the same client-side application code to all users.
Such applications can still fetch individual content (but not code) from the server in later user interactions, though.

\balance
A key premise of \WAIT is the assumption that there is an unbalanced effort for an inside attacker:
secretly deploying an application release is less feasible than simply making changes to the application code.
With \WAIT an attacker must not only re-sign the modified application, but they must also log the release publicly.
Although this process can be automated as part of an continuous integration and continuous deployment pipeline, we suggest that an internal review team should manually approve and sign releases.

\WAIT provides interoperability as browsers without \WAIT support will simply ignore the additional header provided by a \WAIT-enabled web server.
In turn, a capable browser will only apply the integrity protection when the server provides the header.

The append-only guarantees of the transparency log can be\linebreak achieved in several ways:
\emph{Independent log servers}, similar to Certificate Transparency, based on verifiable data structures (\eg Merkle trees) allow validation of the append-only property through constant independent monitoring.
\emph{Trusted computing} (\eg Intel SGX) can provide append-only guarantees by essentially introducing a trusted third party.
A \emph{distributed ledger}\,---\,either through traditional consensus protocols by a fixed set of members (\eg browser vendors) or a proof-of-work-based, publicly verifiable blockchain\,---\,can also provide append-only guarantees.

\bibliographystyle{ACM-Reference-Format}
\bibliography{references}


\begin{thebibliography}{8}

%%% the Web address.

\ifx \showCODEN    \undefined \def \showCODEN     #1{\unskip}     \fi
\ifx \showDOI      \undefined \def \showDOI       #1{#1}\fi
\ifx \showISBNx    \undefined \def \showISBNx     #1{\unskip}     \fi
\ifx \showISBNxiii \undefined \def \showISBNxiii  #1{\unskip}     \fi
\ifx \showISSN     \undefined \def \showISSN      #1{\unskip}     \fi
\ifx \showLCCN     \undefined \def \showLCCN      #1{\unskip}     \fi
\ifx \shownote     \undefined \def \shownote      #1{#1}          \fi
\ifx \showarticletitle \undefined \def \showarticletitle #1{#1}   \fi
\ifx \showURL      \undefined \def \showURL       {\relax}        \fi
% invisible to TeX
\providecommand\bibfield[2]{#2}
\providecommand\bibinfo[2]{#2}
\providecommand\natexlab[1]{#1}
\providecommand\showeprint[2][]{arXiv:#2}

\bibitem[\protect\citeauthoryear{Al-Bassam and Meiklejohn}{Al-Bassam and
  Meiklejohn}{2018}]%
        {al2018contour}
\bibfield{author}{\bibinfo{person}{Mustafa Al-Bassam} {and}
  \bibinfo{person}{Sarah Meiklejohn}.} \bibinfo{year}{2018}\natexlab{}.
\newblock \showarticletitle{Contour: A practical system for binary
  transparency}.
\newblock In \bibinfo{booktitle}{\emph{Data Privacy Management,
  Cryptocurrencies and Blockchain Technology}}. \bibinfo{publisher}{Springer},
  \bibinfo{pages}{94--110}.
\newblock
\urldef\tempurl%
\url{https://doi.org/10.1007/978-3-030-00305-0_8}
\showDOI{\tempurl}


\bibitem[\protect\citeauthoryear{Hodges, Jackson, and Barth}{Hodges
  et~al\mbox{.}}{2012}]%
        {rfc6797}
\bibfield{author}{\bibinfo{person}{Jeff Hodges}, \bibinfo{person}{Collin
  Jackson}, {and} \bibinfo{person}{Adam Barth}.}
  \bibinfo{year}{2012}\natexlab{}.
\newblock \bibinfo{title}{{HTTP Strict Transport Security (HSTS)}}.
\newblock \bibinfo{howpublished}{RFC 6797}.
\newblock
\urldef\tempurl%
\url{https://doi.org/10.17487/RFC6797}
\showDOI{\tempurl}


\bibitem[\protect\citeauthoryear{Laurie, Langley, and Kasper}{Laurie
  et~al\mbox{.}}{2013}]%
        {RFC6962}
\bibfield{author}{\bibinfo{person}{B. Laurie}, \bibinfo{person}{A. Langley},
  {and} \bibinfo{person}{E. Kasper}.} \bibinfo{year}{2013}\natexlab{}.
\newblock \bibinfo{booktitle}{\emph{Certificate Transparency}}.
\newblock \bibinfo{type}{RFC} 6962.
\newblock
\showISSN{2070-1721}


\bibitem[\protect\citeauthoryear{Mozilla}{Mozilla}{2017}]%
        {MozillaBinaryTransparency}
\bibfield{author}{\bibinfo{person}{Mozilla}.} \bibinfo{year}{2017}\natexlab{}.
\newblock \bibinfo{booktitle}{\emph{Security/Binary Transparency}}.
\newblock
\urldef\tempurl%
\url{https://wiki.mozilla.org/Security/Binary_Transparency}
\showURL{%
\tempurl}


\bibitem[\protect\citeauthoryear{Rescorla and Dierks}{Rescorla and
  Dierks}{2008}]%
        {rfc5246}
\bibfield{author}{\bibinfo{person}{Eric Rescorla} {and} \bibinfo{person}{Tim
  Dierks}.} \bibinfo{year}{2008}\natexlab{}.
\newblock \bibinfo{title}{{The Transport Layer Security (TLS) Protocol Version
  1.2}}.
\newblock \bibinfo{howpublished}{RFC 5246}.
\newblock
\urldef\tempurl%
\url{https://doi.org/10.17487/RFC5246}
\showDOI{\tempurl}


\bibitem[\protect\citeauthoryear{{W3C}}{{W3C}}{2016}]%
        {W3SRI}
\bibfield{author}{\bibinfo{person}{{W3C}}.} \bibinfo{year}{2016}\natexlab{}.
\newblock \bibinfo{booktitle}{\emph{Subresource Integrity, W3C
  Recommendation}}.
\newblock
\urldef\tempurl%
\url{https://www.w3.org/TR/2016/REC-SRI-20160623/}
\showURL{%
\tempurl}


\bibitem[\protect\citeauthoryear{{W3C}}{{W3C}}{2017}]%
        {W3CryptoAPI}
\bibfield{author}{\bibinfo{person}{{W3C}}.} \bibinfo{year}{2017}\natexlab{}.
\newblock \bibinfo{booktitle}{\emph{Web Cryptography API, W3C Recommendation}}.
\newblock
\urldef\tempurl%
\url{https://www.w3.org/TR/2017/REC-WebCryptoAPI-20170126/}
\showURL{%
\tempurl}


\bibitem[\protect\citeauthoryear{{W3C}}{{W3C}}{2018}]%
        {W3CSP}
\bibfield{author}{\bibinfo{person}{{W3C}}.} \bibinfo{year}{2018}\natexlab{}.
\newblock \bibinfo{booktitle}{\emph{Content Security Policy, W3C Working
  Draft}}.
\newblock
\urldef\tempurl%
\url{https://www.w3.org/TR/2018/WD-CSP3-20181015/}
\showURL{%
\tempurl}


\end{thebibliography}

\end{document}